\documentclass{llncs}
\usepackage{amsmath,amssymb}
\usepackage{xspace}
\usepackage{comment}
\usepackage{xcolor}
\usepackage{listings}
\usepackage[utf8]{inputenc}
\usepackage[color=yellow!40]{todonotes}
\usepackage{breakcites}
\usepackage{breqn}

\def\isanonymous{0}

\usepackage{ifthen}
\newcommand{\anonymous}[2]{%
\ifthenelse{\equal{\isanonymous}{1}}%
{{#1}}%
{{#2}}%
}

\newcommand{\osstmmquote}[2]{%
  \begin{quote}
  {#1}~{#2}
  \end{quote}
}

\title{The Vacuity of the Open Source Security Testing Methodology Manual}
\anonymous{}{
  \author{Martin R.~Albrecht \and Rikke Bjerg Jensen\institute{Information Security Group, Royal Holloway, University of London}}
}

\begin{document}

\maketitle

\begin{abstract}
  The Open Source Security Testing Methodology Manual (OSSTMM) provides a ``scientific methodology for the accurate characterization of operational security''~\cite[p.13]{Herzog10}. It is extensively referenced in writings aimed at security testing professionals such as textbooks, standards and academic papers. In this work we offer a fundamental critique of OSSTMM and argue that it fails to deliver on its promise of actual security. Our contribution is threefold and builds on a textual critique of this methodology. First, OSSTMM's central principle is that security can be understood as a quantity of which an entity has more or less. We show why this is wrong and how OSSTMM's unified security score, the rav, is an empty abstraction. Second, OSSTMM disregards risk by replacing it with a trust metric which confuses multiple definitions of trust and, as a result, produces a meaningless score. Finally, OSSTMM has been hailed for its attention to human security. Yet it understands all human agency as a security threat that needs to be constantly monitored and controlled. Thus, we argue that OSSTMM is neither fit for purpose nor can it be salvaged, and it should be abandoned by security professionals.
\end{abstract}

\section{Introduction}\label{sec:introduction}

Penetration testing textbooks advise their readers to follow a pre-established methodology. For example, Johansen et al.~write: ``A penetration testing methodology defines a roadmap, with practical ideas and proven practices that can be followed to assess the true security posture of a network, application, system, or any combination thereof''~\cite{JAHA16}. Similarly, Duffy notes: ``The biggest benefit of using a methodology is that it allows assessors to evaluate an environment holistically and consistently'' and ``when standard exploits do not work, testers can have tunnel vision; sticking to a methodology will prevent that''~\cite[pp.5-6]{Duffy15}. Penetration testing methodologies are therefore seen to enable a systematic assessment of an organisation's security.

However, the use of a penetration testing methodology contains within it a tension. On the one hand, it ought to provide a complete coverage of the target, thus enabling a better understanding of its security; deciding on a methodology before engagement should enable a better understanding of the object afterwards. On the other hand, fixing the steps and tests performed to understand the security of a target before engaging with it, may subvert the understanding of it. The methodology may simply not be adequate for the object under consideration. For example, if a methodology does not cover IPv6, attack vectors via IPv6 will be missed unless the tester deviates from the methodology under their own initiative. Similarly, vulnerabilities involving, say, SCTP traffic are unlikely to be captured, since methodologies typically focus on TCP and UDP\@.

This tension is, for example, identified by Wilhelm when he writes: ``What we need in our industry is a repeatable process that allows for verifiable findings, but which also allows for a high degree of flexibility on the part of the pentest analyst to perform `outside-the-box' attacks and inquiries against the target systems and networks''~\cite[p.76]{Wilhelm13}. Similarly, Stuttard and Pinto emphasise: ``Following all the steps in this methodology will not guarantee that you discover all the vulnerabilities within a given application. However, it will provide you with a good level of assurance that you have probed all the necessary regions of the application's attack surface and have found as many issues as possible given the resources available to you''~\cite{StuPin11}.

This tension does not invalidate the utility of penetration testing methodologies in many scenarios as the tested objects tend to exhibit a large level of similarity, permitting presumptions to be made about the objects under consideration. It does, however, caution against claims of actual security, i.e.~a full understanding of the object under consideration, when the object was not, in fact, studied in its own right but through the lens of a predecided methodology. The standardised nature of such methodologies also questions their ability to yield reliable results about an organisation's total security posture. 

Significantly, however, this limitation of penetration testing methodologies is not necessarily acknowledged by the methodologies themselves. In particular, the Open Source Security Testing Methodology Manual (OSSTMM), which we consider in this work, promises an accurate understanding of security -- what it terms ``Actual Security'' -- as the result of the application of its scientific methodology: 

\begin{quote}
The primary purpose of this manual is to provide a scientific methodology for the accurate characterization of operational security (OpSec) through examination and correlation of test results in a consistent and reliable way.~\cite[Introduction,~p.13]{Herzog10}
\end{quote}

\subsubsection{OSSTMM.} The methodology was first introduced in 2000. The current version is 3.0 and was released in 2010 by the Institute for Security and Open Methodologies (ISECOM). There is also a draft version 4.0, but it seems to be hardly considered, plausibly due to the fact that it is only available to ISECOM members.

OSSTMM is structured similarly to other security testing methodologies. It introduces its basic premises, notions and processes in chapters one to six. This is followed by five chapters on particular areas, each discussing concrete tests. The methodology finishes with pointers on compliance, reporting, expected outcomes and the licence. OSSTMM opens its Introduction with:
 
\begin{quote}
The Open Source Security Testing Methodology Manual (OSSTMM) provides a methodology for a thorough security test, herein referred to as an OSSTMM audit. An OSSTMM audit is an accurate \emph{measurement} of security at an operational level that is \emph{void of assumptions} and anecdotal evidence.~\cite[Introduction,~p.11, emphasis added]{Herzog10}
\end{quote}

With this, the authors announce OSSTMM's two out of three core contributions to security testing which distinguish it from other methodologies: its security metrics (Chapter~4) and its trust analysis (Chapter~5).

First, OSSTMM defines a unified security \emph{score} -- the rav -- to be measured which ought to express the deviation from perfect security: 

\begin{quote}
The rav is a scale measurement of an attack surface, the amount of uncontrolled interactions with a target, which is calculated by the quantitative balance between porosity, limitations, and controls. In this scale, 100 rav (also sometimes shown as 100\% rav) is perfect balance and anything less is too few controls and therefore a greater attack surface. More than 100 rav shows more controls than are necessary which itself may be a problem as controls often add interactions within a scope as well as complexity and maintenance issues.~\cite[Ch.1,~p.22]{Herzog10}
\end{quote}

Second, for the avoidance of bias and reliance on assumptions, OSSTMM defines security independent of risk, the environment and threats: 

\begin{quote}
However, to remove bias from security metrics and provide a more fair assessment we removed the use of risk. Risk itself is heavily biased and often highly variable depending on the environment, assets, threats, and many more factors.~\cite[Ch.1,~p.28]{Herzog10}
\end{quote} 

\noindent To avoid the pitfalls it associates with risk, OSSTMM proposes quantifiable, fact based \emph{trust} metrics: 

\begin{quote}
Our intention is to eventually eliminate the use of risk in areas of security which have no set price value of an asset (like with people, personal privacy, and even fluctuating markets) in favor of trust metrics which are based completely on facts.~\cite[Instructions,~p.2]{Herzog10}
\end{quote}

The third major contribution of OSSTMM is its ``holistic''~\cite[Ch.4,~p.68]{Herzog10} approach to security. That is, OSSTMM applies this metric and its methodology to a comprehensive variety of areas, including, and in contrast to other such methodologies, to \emph{Human Security} (Chapter~7): 

\begin{quote}
This is a methodology to test the operational security of physical locations, human interactions, and all forms of communications such as wireless, wired, analog, and digital.~\cite[Instructions,~p.2]{Herzog10}
\end{quote}

This contribution is often highlighted in the literature, in e.g.~\cite{PraRam10} OSSTMM is recognised for being the first methodology ``to include human factors in the tests, taking into account the established fact that humans may be very dangerous for the system''.\footnote{The seminal work criticising this notion is~\cite{AdaSas99}, see Section~\ref{sec:related-work}.}

\subsubsection{OSSTMM's Impact.} ISECOM offers various certifications for security professionals, such as OPST (OSSTMM Professional Security Tester), OPSE (OSSTMM Professional Security Expert) and CTA (Certified Trust Analyst), and for organisations, infrastructure and products, such as STAR (Security Test Audit Report) and the OSSTMM Seal of Approval. Furthermore, ISECOM has several related projects that build on OSSTMM, such as SCARE (Source Code Analysis Risk Evaluation) which applies the rav to source code analysis, HSM (Home Security Methodology and Vacation Guide) which applies the rav to securing a home, HHS (Hacker Highschool) which teaches security awareness to teenagers based on OSSTMM, and BPP (The Bad People Project) which is a security and safety awareness programme for children and parents based on OSSTMM's rav and trust metrics. These projects further emphasise the centrality of the rav, trust and human interactions, i.e.~the aspects of OSSTMM focused on in this work, to the ISECOM mission. 

Beyond these affiliated projects, it is difficult to assess how widely OSSTMM is used. However, in a 2015 survey~\cite{KnoBarMcg15}, 10 out of 32 penetration testing providers cite OSSTMM as an influence for their own methodology.\footnote{For context, 16 mentioned OWASP generally but only three mentioned the OWASP testing guide, PTES was mentioned by six providers, three providers mentioned NIST SP 800-115.}
Furthermore, CREST's \emph{A guide for running an effective Penetration Testing programme} refers to OSSTMM as an authoritative source for a ``standard penetration testing methodolog[y]'' and notes its comprehensiveness~\cite{CREST:Guide17}. NIST Special Publication 800-115 calls it a ``widely used assessment methodology''~\cite{NIST:SP800-115} and the PCI Penetration Testing Guidance for the PCI Data Security Standard (PCI DSS) lists it as one of five ``industry-accepted methodologies''~\cite{Council15}. A similar note is struck in~\cite{Shackleford14}: ``There are not many standards in use today for assessments and pen tests: PTES, OSSTMM, etc.''\footnote{See Section~\ref{sec:related-work} for a discussion of other security testing methodologies.}; in~\cite{HHMNZ14} Holik et al.~refer to OSSTMM as ``heavily reputable in penetration testers community'';in~\cite{PraRam10} Prandini and Ramilli go further and refer to OSSTMM as ``the de-facto standard for security testers'' and in~\cite{FMLPMCSPSP14} the authors refer to it as one of ``the two most important standards in cyber-security''. OSSTMM is referenced in many textbooks on penetration testing, e.g.~\cite{Wilhelm13,OS14,Duffy15,JAHA16,Mcphee17}, is the methodology of choice in ``Hacking Exposed Linux''~\cite{ISECOM08}, and is used in~\cite{Schulte09} for Information Assurance testing and certification by the US Defense Information Systems Agency (DISA). Several academic works reference and build on OSSTMM, e.g.~\cite{CZCG05,CCZC05,FMPLM15,Jimenez16,CasKar16,TFSJD18}. Overall, OSSTMM has more than 150 citations according to Google Scholar as of Summer 2020. 

\subsubsection{Contributions.} In this work, we investigate if the Open Source Security Testing Methodology Manual delivers on what it promises with a particular focus on its main contributions. As a consequence, we offer a fundamental critique of this methodology. We do so by addressing its three main contributions, OSSTMM's raisons d'être, in turn.

First, OSSTMM's central premise and promise is that security across many areas can be understood as a \emph{quantity} of which an entity has more or less. In Section~\ref{sec:the-rav}, we explain and demonstrate why this is incorrect and show how OSSTMM fails to deliver on its promise of bringing security to the fore by supplanting an understanding of its object with a method that disregards it. The end result of this process, the rav, is a number that can be readily calculated but conveys little about the security of the considered object.

Second, OSSTMM's Trust definition, which is essential to the methodology and intended to replace risk, not only shares the shortcomings of the rav, but also collapses under its own contradictions on inspection. In particular, OSSTMM's attempt to identify sociological, psychological and technical notions of trust produces nonsensical claims and meaningless metrics, as we explain in Section~\ref{sec:value-of-trust}. 

Third, this leads us to question OSSTMM's decision to treat all areas of security the same in Section~\ref{sec:human-security}. This decision, combined with the resolve to disregard any notion of risk or threat, leads OSSTMM to conceptualise \emph{all} human agency as a security threat. This results in a Human Security testing approach treating all employees effectively as potential insurgents and testing procedures that test the ``requirements to incite fear, revolt, violence, and chaos''~\cite[Ch.7,~p.110]{Herzog10}.

Our critique of OSSTMM's core components invalidates its intended function. Overall, we find that OSSTMM imposes its abstractions against the reality they are designed to model, producing an understanding of security that is empty at best and outright harmful when Human Security is concerned. As such, our conclusion is that OSSTMM's approach to security (testing) cannot be salvaged and the use of OSSTMM should be abandoned by practitioners. We discuss this further and set out broader lessons in Section~\ref{sec:conclusion}. 

\subsubsection*{Method.}\label{sec:nature-this-work} While information security research routinely features critiques of security technologies in the form of ``attack papers'', analogues of such works for policies, frameworks and conceptions are largely absent from its core venues. This work is a textual critique of OSSTMM based on a close reading of the methodology and pursues two purposes. First, immediately, to show that OSSTMM is inadequate as a security testing methodology, despite being referenced routinely in the security testing literature. Second, more mediated, to show that the \emph{ideas} at the core of OSSTMM are wrong. As we show in Section~\ref{sec:conclusion}, these ideas are not OSSTMM's privilege. It is for this reason that we chose the form of a textual critique over alternative approaches such as empirical studies to the effectiveness of OSSTMM in practice.

\section{The Rav}\label{sec:the-rav}

The central concept in OSSTMM is its security score -- the \emph{rav} -- as illustrated by ISECOM's ``OSSTMM Seal of Approval'': 

\begin{quote}
This seal defines an operational state of security, safety, trust, and privacy. The successfully evaluated products, services, and processes carry their visible certification seal and rav score. This allows a purchaser to see precisely the amount and type of change in security that the evaluated solutions present. It removes the guesswork from procurement and allows one to find and compare alternative solutions.~\cite[Introduction,~p.16]{Herzog10}
\end{quote}

\subsection{Health Analogy}\label{sec:health-analogy}

OSSTMM illustrates its approach with an analogy to health in its chapter on Operational Security Metrics. Since this analogy provides a succinct summary, we also start there. 

\begin{quote}
Imagine a machine exists that can audit all the cells in a human body. This machine works by monitoring the cells in their environment and even prodding each cell in a way it can react to better categorize its purpose. We could then see what various cells do and how they contribute to the overall make-up of the human body. Some cells make up tissue walls like skin cells do. Some, like white blood cells, provide authentication and attack other cells which are on its ``bad'' list. Then some cells are foreigners, like bacteria which have entered at some point and thrived. The machine would classify all the cells that make up the person, a defined scope, rather than say which are ``bad'' or ``good''.~\cite[Ch.4,~p.64]{Herzog10}
\end{quote}

The starting point of the thought experiment is the ability to audit every cell in a human body. While the analogy appeals to an atomic view of health -- cells -- initially the audit is not atomic. That is, the hypothetical machine observes cells ``in their environment'' as part of the human body. Yet, this perspective is immediately abandoned: 

\begin{quote}
By counting the cells the machine can tell mostly how well the person as an organism works (health) and how well they fit into their current environment. It can also determine which cells are broken, which are superfluous, and of which type more might be required for the person to be more efficient, prepared for the unexpected, or for any number of specific requirements. Since the cells are dividing and dying all the time, the machine must also make regular tests and chart the person’s ability to improve or at least maintain homeostasis.~\cite[Ch.4,~p.64]{Herzog10}
\end{quote}

OSSTMM moves from understanding cells in their environment to \emph{counting} them. This transition is premised on the fact that these different parts of the body are all called ``cells'' which gives the impression that they share the same unit and the premise that they can be \emph{added up} to understand a person's health. Furthermore, this is premised on the idea that health is a \emph{totally ordered set}. Finally, note that OSSTMM here promises to make ``the person [\ldots] more efficient [\ldots] for any number of specific requirements'' by exclusively considering the inner workings of the body without any reference to for what \emph{purpose} said person ought to be made more efficient. We shall see below that these ideas shape OSSTMM's consideration of security.

\subsection{Introducing the Rav}\label{sec:introducing-rav}

Having introduced the key ideas of its security metric, OSSTMM is ready to introduce its security score: the rav\@. 

\begin{quote}
Unfortunately there is no such machine for counting all cells in a human body. However it does exist for security. Analysts can count and verify the operations of targets in a scope as if it is a super-organism. They record its interactions and the controls surrounding those interactions. They classify them by operation, resources, processes, and limitations. Those numbers the Analysts generate are combined so that controls add to operational security and limitations take away from it. Even the value of the limitations, how badly each type of problem hurts, is also not arbitrary because it’s based on the combination of security and controls within that particular scope. So a bad problem in a protective environment will provide less overall exposure than one in a less controlled environment.~\cite[Ch.4,~p.64]{Herzog10}
\end{quote}

As it does for health, OSSTMM considers security -- Actual Security in OSSTMM terms -- to be \emph{one} ranking in which an entity scores higher or lower, a totally ordered set. It is, perhaps, common to say ``System A is more secure than System B because System A has Advanced Cyber Thread Analysis Blockchain Technology™''. However, is a system with a local privilege escalation bug that requires physical access more or less secure \emph{per se} than a server vulnerable to a Denial of Service attack from an IoT botnet? Is a person using a menstruation app to avoid pregnancy that is running on a phone to which their partner has access more or less secure than a server running the stable release of Debian GNU/Linux patched four days ago? These questions make no sense. What we mean by security depends on the object and the threats we are considering.\footnote{Indeed, in e.g.~cryptography where quantitative statements of security are abound in the form of advantages and computational complexities, these are always related to specific security goals and attacker capabilities. For example, any cryptographic textbook will distinguish between the collision resistance and preimage resistance of a cryptographic hash function and will shy away from unifying those into one score.}

OSSTMM computes its rav score from ``Porosity'' (also referred to as ``OpSec''), ``Controls'' and ``Limitations''.

\subsubsection{Porosity.}\label{sec:porosity} To establish Actual Security, OSSTMM starts by considering the porosity of the target, where each ``pore'' is either ``Visibility'', ``Access'', or ``Trust''. For example, a server with ports 22, 80 and 443 open, would have an Access of three. Since porosity is always considered as a negative for security, the minimal rav in this example would be -3 (up to some normalisation). 

\begin{quote}
The minimum rav is made by the calculation of porosity which are the holes in the scope. The problem with security metrics is generally in the determination of the assessors to count what they can't possibly really know. This problem does not exist in the rav. You get what you know from what is there for a particular vector and you make no assumptions surrounding what is not there. You count all that which is visible and interactive outside of the scope and allows for unauthenticated interaction between other targets in the scope. That becomes the porosity. This porosity value makes the first of 3 parts of the final rav value.~\cite[Ch.4,~p.67]{Herzog10}
\end{quote}

We will return to porosity, with a focus on Trust, in Section~\ref{sec:value-of-trust}. For now, note that OSSTMM's critique of other security metrics is that they aim to count what they cannot possibly know. That indeed sounds like something worth avoiding. Presumably, though, those metrics wish to include a certain bit of information -- which they nevertheless do not have -- because it is \emph{relevant}. Thus, there is a dilemma: we need information which we do not have. OSSTMM resolves this dilemma by discarding what it needs to know in order to simply count what is known. The task -- understanding the security of the object at hand -- is replaced by a counting method whose appeal is merely that it is \emph{feasible}. It is one thing to give an account of what you know, it is another thing entirely to claim that whatever you are able to observe from your vantage point is the correct understanding of the object, when you \emph{know} it is not.

\subsubsection{Controls.}\label{sec:controls} Next, OSSTMM identifies control classes, all of which are always to be acknowledged in all domains, or ``channels'', that OSSTMM considers: Human, Physical, Wireless, Telecommunications, and Data Network Security. Thus, from the perspective of OSSTMM, these different domains share a high level of similarity. Here, the argument relies on homographs. We illustrate this using the control class ``Integrity'':

\begin{quote}
Count each instance for Access or Trust in the scope which can assure that the interaction process and Access to assets has finality and cannot be corrupted, stopped, continued, redirected, or reversed without it being known to the parties involved. Integrity is a change control process.
In COMSEC data networks, encryption or a file hash can provide the Integrity control over the change of the file in transit.\footnote{It is worth noting that neither cryptographic mechanism described here provides integrity protection: for example, CBC mode encryption and textbook RSA are famously malleable, e.g.~\cite{SP:AlfPat13,C:Bleichenbacher98}, and hash functions are public functions operating on public data so an adversary can simply recompute the hash after message modification. The authors should have recommended a message authentication code or a digital signature.} In HUMSEC, segregation of duties and other corruption-reduction mechanisms provide Integrity control. Assuring integrity in personnel requires that two or more people are required for a single process to assure oversight of that process. This includes that no master Access to the whole process exists. There can be no person with full access and no master key to all doors.~\cite[Ch.4,~p.70]{Herzog10}
\end{quote}

The COMSEC example in the above quote refers to the integrity of messages, i.e.~the prevention of message modification by someone other than the sender. The recommended controls are intended to ensure that whatever message the sender intended to send is indeed received. The HUMSEC example, however, is concerned with distrust in the sender. It recommends ``corruption-reduction mechanisms'' to hedge against the originators of actions. Thus, the two controls are aimed at different threats: the first aims to ensure an honest actor's messages are not corrupted in flight, the second aims to ensure an actor itself is not corrupt.\footnote{An analogous cryptographic control would be the use of secret sharing and secure multiparty computation techniques.} The only relation is that the words ``integrity'' and ``corruption'' are used in both cases. We see that the identification is merely facilitated by their identical linguistic features and not by their distinct meanings. The control classes identified by OSSTMM are anything but self-evident. Yet, OSSTMM does not justify them.

Overall OSSTMM defines ten such control classes -- ``authentication'', ``indemnification'', ``subjugation'', ``continuity'', ``resilience'', ``non-repudiation'', ``confidentiality'', ``privacy'', ``integrity'', and ``al\-arm''~\cite[p.72-75]{Herzog10}. The controls from all the classes are then \emph{added up} with weights 1/10 for each class. 

\begin{quote}
The next part is to account for the controls in place per target. This means going target by target and determining where any of the 10 controls are in place such as Authentication, Subjugation, Non-repudiation, etc. Each control is valued as 10\% of a pore since each provides 1/10th of the total controls needed to prevent all attack types. This is because having all 10 controls for each pore is functionally the same as closing the pore provided the controls have no limitations.~\cite[Ch.4,~p.67]{Herzog10}
\end{quote}

For example,\footnote{We give a full worked example of a rav calculation in Appendix~\ref{sec:toy-example}.} two such controls would be ``log file is in place'' and ``authentication is required'' and we would obtain \[1/10 \times \textnormal{``log file is in place''} + 1/10 \times \textnormal{``authentication is required''}.\] At this point, the reader may think of the above expression as a formulaic way of saying ``one log file is in place and authentication is required''. However, as OSSTMM explains above, this is meant to be a weighted sum where each summand is ``valued'' at 1/10, i.e.~as far as OSSTMM is concerned, the sum of a log file and ``authentication required'' is meaningful in a mathematical sense. This reasoning presumes that all controls are the same in some quantifiable way. The authors of OSSTMM explain: 

\begin{quote}
It is difficult to work with relative or inconsistent measurements like choosing a specific hue of yellow to paint a room, starting work at sunrise, having the right flavor of strawberry for a milkshake, or preparing for the next threat to affect your organization’s profits because the factors have many variables which are biased or frequently changing between people, regions, customs, and locations. For this reason, many professions attempt to standardize such things like flavors, colors, and work hours. This is done through reductionism, a process of finding the elements of such things and building them up from there by quantifying those elements. This way, colors become frequencies, work hours become hours and minutes, flavors become chemical compounds, and an attack surface becomes porosity, controls, and limitations. The only real problem with operational metrics is the requirement for knowing how to properly apply the metric for it to be useful.~\cite[Ch.4,~p.62]{Herzog10}
\end{quote}

OSSTMM claims to have identified ``the elements'' of a log file and the requirement for authentication. However, while e.g.~adding up the frequencies of colours indeed produces a new colour (frequency), i.e.~colours have a quantitative side to them that permits addition, this does not hold true for log files and authentication requirements.

One log file and one authentication requirement is not the same as one log file and a message authentication code; log files, message authentication codes and (user) authentication requirements are different and protect against different threats. Thus, there are no inherent weights when adding these things that are not reducible to the same dimension.

Even assuming that ``having all 10 controls for each pore is functionally the same as closing the pore provided the controls have no limitations'', this does not imply adding up (a subset of) ten controls to obtain a score since each of the specific controls needs to be in place. OSSTMM's appeal
to the diversity of the controls does not produce the posited identity. Rather, the sentence merely implies that the ten classes should sum to one (or whatever stands in for ``all good'') when they are all in place. The weights are irrelevant when all control classes are in place since they are
designed to sum to one in this case, but matter for when a particular control class is missing, i.e.~they are meant to encode the importance attributed to this particular lack. This, in turn, will depend on what is being protected and the nature of the threat under consideration. It is not at all clear that adding an authentication requirement closes a pore to the same extent as adding a log file documenting access after the fact does. The \emph{choice} 1/10 ``authentication required'' and 1/10 ``log file'' is as much a choice as \(\sqrt{2\pi/e}\) ``authentication required'' and 1/10
``log file''. OSSTMM is not ``void of assumptions''~\cite[Introduction,~p.11]{Herzog10}.

\subsubsection{Limitations.}\label{sec:limitations} To complete the rav, limitations are considered:

\begin{quote}
The third part of the rav is accounting for the limitations found in the protection and the controls. These are also known as ``vulnerabilities''. The value of these limitations comes from the porosity and established controls themselves.~\cite[Ch.4,~p.67]{Herzog10}
\end{quote} 

To make this concrete, too, we may think of this operation of adding up several ``buffer overflows'' with, say, one ``CR/LF escape''. 

\subsubsection{Sums.}\label{sec:sums} With all components in place, the rav can be calculated.

\begin{quote}
With all counts completed, the rav is basically subtracting porosity and limitations from the controls. This is most easily done with the rav spreadsheet calculator.~\cite[Ch.4,~p.67]{Herzog10}
\end{quote}

Thus, OSSTMM goes beyond the idea of security as a ranking and considers security to possess an additive structure.\footnote{These are not equivalent. The letters in the alphabet are ordered but this does not endow them with an addition rule. The integers modulo some prime \(p\) have an addition (multiplication, division) rule but do not possess a natural ordering.} Enabling encryption ``adds to'' security, having a telnet port open permitting root login ``takes away'' security.\footnote{``Those numbers the Analysts generate are combined so that controls add to operational security and limitations take away from it''~\cite[Ch.4,~p.64]{Herzog10}} Mathematics also speaks of ``adding'' and ``taking away'', when referring to addition and subtraction over, say, the Integers, and \emph{thus}, so OSSTMM's leap, we shall add and subtract open ports and authentication:
\[c_{0} \times \textnormal{log file} + c_{1} \times \textnormal{auth.} - c_{2} \times \textnormal{open port} - c_{3} \times \textnormal{buf. overfl.},\] where \(c_{i}\) are some weights.\footnote{These weights are not always constants. In particular, limitations are weighted according to their class and porosity. Thus, the actual expression is more complex than given here for illustration purposes.} These sums can further be extended to not only cover the moments of an IT system but across all areas considered by OSSTMM\@: 

\begin{quote}
One important requirement in applying the rav is that Actual Security can only be calculated per scope. A change in channel, vector, or index is a new scope and a new calculation for Actual Security. However, once calculated, multiple scopes can be combined together in aggregate to create one Actual Security that represents a fuller vision of the operational security [of] all scopes.
For example, a test can be made of Internet-facing servers from both the Internet side and from within the perimeter network where the servers reside. That is 2 vectors. Assume that, the Internet vector is indexed by IP address and contains 50 targets. The intranet vector is indexed by MAC address and is made of 100 targets because less controls exist internally to allow for more collaborative interaction between systems. Once each test is completed and the rav is counted they can be combined into one calculation of 150 targets as well as the sums of each limitations and controls. This will give a final Actual Security metric which is more complete for that perimeter network than either test would provide alone. It would also be possible to add the analysis from physical security, wireless, telecommunications, and human security tests in the same way. Such combinations are possible to create a better understanding of the total security in a holistic way.~\cite[Ch.4,~p.68]{Herzog10}
\end{quote}

In other words, OSSTMM not only adds and subtracts ports and log files but also doors (Physical Security), ``whispering or using hand signals''~\cite[Ch4.,~p.74]{Herzog10} or ``a cultural bias''~\cite[Ch.4,~p.76]{Herzog10} (Human Security). The computation continues for a few more steps to produce the final rav value (see Appendix~\ref{sec:toy-example} for a more detailed example). However, already at this stage this score is an arbitrary choice produced by eradicating the differences between the features considered in order to combine them using some chosen weights. As its opening gambit OSSTMM promises the reader the avoidance of ``general best practices, anecdotal evidence, or superstitions''~\cite[p.1]{Herzog10}, but the answer OSSTMM gives as to the ``Actual Security'' of the studied object is vacuous and is based on category mistakes and unsubstantiated choices. 

\begin{remark}
While it is possible to construct examples where an OSSTMM score contradicts an expert's verdict on security, since e.g.~all controls are weighted the same, being vacuous does not mean that the score must commonly and strikingly disagree with the reality it subsumes. Consider the following hypothetical example: if the rav score for, say, an unpatched Windows 7 system exposing SMB on the Internet were to be lower than for, say, an up-to-date copy of OpenBSD in the default configuration, no one would question the security of the latter in favour of the former. Rather, if this was the case, the rav would change. In contrast, if we were to rethink the security of a patched OpenBSD and an unpatched Windows 7, it would be because we had learned something new about these systems. Echoing Kay~\cite{Kay09}, the rav tells us nothing that we have not previously told the rav. It does, however, obscure what we tell it. The rav score itself is not useful to an engineer tasked with improving it: the engineer would have to return to the data that was used to compute it to understand which security controls were missing and where. The sentence ``telnet is open'' contains more information than whatever numerical value OSSTMM assigns to it.
\end{remark}

\section{Trust}\label{sec:value-of-trust}

As mentioned above, in OSSTMM terms, Access, Visibility and Trust make up Porosity (also referred to as OpSec), and security is defined as the separation of a threat and an asset. While Access is roughly what you would expect -- the ability to interact with the asset -- Visibility is ``a means of calculating opportunity''~\cite[Introduction,~p.23]{Herzog10}. From this perspective, an asset needs to be visible to be targeted. 

For OSSTMM, Trust is a core component of security testing and, unlike e.g.~Access and Visibility, requires its own separate analysis, which is why OSSTMM dedicates the whole of Chapter~5 to it. This attention to Trust is chiefly motivated by ISECOM's ambition to replace considerations of risk with trust metrics, as discussed in our introduction. Thus, examining Trust in OSSTMM is examining one of its key tenets. The centrality of this notion to ISECOM's mission is underlined by ISECOM offering certification specifically for its Trust analysis:

\begin{quote}
The Certified Trust Analyst proves a candidate has the skills and knowledge to efficiently evaluate the trust properties\footnote{We discuss OSSTMM's ten trust properties in Section~\ref{sec:properties}.} of any person, place, thing, system, or process and make accurate and efficient trust decisions.~\cite[Introduction,~p.15]{Herzog10}
\end{quote}

Whether applied to a person, an object or a process, Trust has a numerical value -- zero or more Trusts -- that determines ``Trust'', the unit of analysis.

\subsection{What Trust?}\label{sec:what-trust}

In what follows, we use ``Trust'' to denote the concept of trust as defined in OSSTMM and ``trust'' for its wider conceptions, such as the diverging notions of trust in computer science and the social sciences, as outlined in our discussion of related work in Section~\ref{sec:related-work}. Critically, this distinction and the plurality of trust definitions are not recognised by OSSTMM\@. Rather, in order for its trust analysis to capture what it needs it to capture, OSSTMM collapses multiple trust definitions, making its understanding and application of trust nonsensical. While OSSTMM maintains that ``people [\ldots] misuse trust as a concept''~\cite[Ch.5,~p.87]{Herzog10}, we show how this statement directly applies to OSSTMM. We identify and discuss this below by showing how OSSTMM needs to appeal to understandings of trust within computer science as well as sociological and behavioural notions of trust in order to facilitate its trust analysis.

\subsubsection{Computer Science.} The Trust unit in OSSTMM is better understood as modelling the need to trust. Trust is always measured as a negative in OSSTMM, a person with a Trust score of five, for example, is understood to be a riskier proposition than someone with a Trust/need-to-trust score of, say, two (the same goes for any other object):
 
\begin{quote}
Where security is like a wall that separates threats from assets, trust is a hole in that wall.~\cite[Ch.5,~p.87]{Herzog10}
\end{quote}

Trust in OSSTMM terms is couched in an understanding of trust as a risk of exploitation and as a vulnerability. This notion of trust is not necessarily controversial from a computer science perspective. For example, this mirrors the notion of trust in cryptography where constructions not relying on a trusted third party are considered preferable to ones that do. Indeed, cryptographic protocols can be characterised as emulating a trusted third party by mutually distrustful parties.\footnote{As Goldreich writes: ``A general framework for casting (\(m\)-party) cryptographic (protocol) problems consists of specifying a random process that maps \(m\) inputs to \(m\) outputs. The inputs to the process are to be thought of as local inputs of \(m\) parties, and the \(m\) outputs are their corresponding (desired) local outputs. The random process describes the desired functionality. That is, if the \(m\) parties were to trust each other (or trust some external party), then they could each send their local input to the trusted party, who would compute the outcome of the process and send to each party the corresponding output. A pivotal question in the area of cryptographic protocols is the extent to which this (imaginary) trusted party can be `emulated' by the mutually distrustful parties themselves.''~\cite{Goldreich04}}

\subsubsection*{Sociological.} More specifically, OSSTMM defines Trust as Internal Access.

\begin{quote}
In operational security, Trust is merely a contributor to porosity, just another interaction to control. It differs from Access (the other form of interaction), in how it relates to other targets within the scope. So where Access is interaction between two sides of a vector into and out of the scope, Trust is measured as the interactions between targets within the scope.~\cite[Ch.5.~p.87]{Herzog10}
\end{quote}

This notion of trust assumes interaction and thus relies on relations between objects within the environment. Recalling that OSSTMM also attempts to model Human Security, it thus models human interactions which are social in nature. Put differently, access -- internal or not -- is (also) social. Thus, we observe that OSSTMM relies on a sociological conception of trust precisely because it insists that Trust is understood as Internal Access.

However, trust, regardless of disciplinary grounding, is fundamentally different from access and it needs to be recognised as such.\footnote{For example, the cryptographic literature would not refer to interactions, typically modelled as oracle access, in a cryptographic protocol as ``trust''.} It is therefore not only unhelpful to employ the two terms in similar ways, as done here by OSSTMM, it also obfuscates both the cognitive and emotional aspects that make up trust interactions~\cite{LewWei85}. Furthermore, OSSTMM's Trust metric assumes that trust relations do not exist outside the environment as this is simply considered as Access by OSSTMM\@. Thus, why label Internal Access as Trust? If Trust is simply another form of Access, shaped by different types of interactions, why not declare it as such?

\subsubsection*{Behavioural.} On the same page, OSSTMM introduces a third definition:

\begin{quote}
Trust is a decision. While some people claim it is an emotion, like love, or a feeling, like pain, its clearly a complex quality we humans are born with. Unlike an emotion or a feeling, we can choose to trust or not to trust someone or something even if it feels wrong to do so. It appears that we are capable to rationalize in a way to supersede how we feel about trusting a target.~\cite[Ch.5,~p.87]{Herzog10}
\end{quote}

Here, OSSTMM posits trust as either an emotion/feeling or a decision, in order to then reject the former in favour of the latter.\footnote{We may note a category mistake in the initial dichotomy being offered: trust is a content of thought -- what we think -- whereas an emotion is a form -- how it appears. However, since this line of enquiry would take us away from the object at hand, we abandon it here.} OSSTMM is interested in this conclusion because it wants to posit that trust is an object of reason or rationality: trust is open to reflection.\footnote{This is why OSSTMM can afford to contradict itself by re-admitting trust emotions one sentence later. It is not actually invested in the either-or question but that the former can override the latter.} This is not controversial. However, declaring trust as an atomic decision is. Since OSSTMM insists that Trust is Internal Access, and given its appeal to Human Security, it requires a broader understanding of trust; an understanding that acknowledges the contextual and the ``environment'' in which trust interactions emerge and take form. Within these environments, trust does not exist in isolation, in the same way that trust-decisions are not made in a vacuum. Rather, here, trust is a key building block of the (social) environment in which interactions take place -- hence, OSSTMM's notion that trust is~\emph{solely} a choice is deceptive in this context.  

This is also evident in the wider social sciences where, for example, in psychological terms, a person's self-efficacy is critical to any understanding of how an individual makes security decisions~\cite{AMP:Ban82}. From a sociological perspective, societal structures and interpersonal relations influence how people make security decisions~\cite{WebCar03}. From an organisational perspective, workplace culture as well as formal and informal policies influence security decisions~\cite{CoS:VNVS10,IMC:DaGr14}. Exemplified by these works, trust decisions, like security, are deeply interwoven into human relations and contextual settings.

\subsection{Trust Properties}\label{sec:trust-properties}

\subsubsection{Countable.}\label{sec:countable} However, these objections are moot when we recall that OSSTMM simply defined Trust as Internal Access. Yet, so is OSSTMM's discussion above: ``While some people claim \emph{internal access} is an emotion, like love, or a feeling, like pain, it is clearly a complex quality we humans are born with'' makes no sense. OSSTMM wants both: to redefine Trust as Internal Access \emph{as well as} maintaining this redefinition captures the notion of trust as a human capacity. This confusion is meant to enable OSSTMM to make trust quantifiable for the rav: 

\begin{quote}
This means we can quantify it by applying a logical process. It also means we can assign trust values to objects and processes as well as people based on these values. This brings new power to those who can analyze trust and make decisions based on that analysis.~\cite[Ch.5,~p.87]{Herzog10}
\end{quote}

This does not mean that at all. As far as OSSTMM is concerned, to rationally understand an object, to reason about it, means to quantify it; a leap OSSTMM simply asserts here. This is like saying to reason about OSSTMM we should count, say, its number of pages, words, characters, revisions and so on. Or, since OSSTMM appeals to mathematics, this is like saying we understand the ring \(\mathbb{Z}_{7681}[x]/(x^{256} + 1)\) when we know it has \(7681^{256}\) elements.

\subsubsection{Properties.}\label{sec:properties} To compute Trust, OSSTMM proposes ten Trust Properties~\cite[Ch.5,~p.90]{Herzog10}. These range from seemingly calculable notions such as size and value to more evasive ones such as symmetry and consistency:\footnote{The ten Trust Properties identified by OSSTMM are: Size, Symmetry, Visibility, Subjugation, Consistency, Integrity, Offsets, Value, Components, Porosity.} 

\begin{quote}
The Trust Properties are the quantifiable, objective elements which are used to create trust. We can say these properties are what we would say give us ``reason to trust''.\footnote{As we shall see below, these are actually reasons \emph{not} to trust, i.e.~a high score in one of the Trust Properties implies a high need to trust.} These properties are to be made into baseline rules based on the target and situation which we are verifying. During research, many potential Trust Properties were discovered which are commonly in use and even official, government and industry regulations recommend [sic.], however they failed logic tests and were discarded from our set of properties leaving only ten.~\cite[Ch.5,~p.88]{Herzog10}
\end{quote}

As this paragraph exemplifies, all Trust Properties identified in the methodology are treated as quantifiable and unbiased. However, at no point does OSSTMM attempt to explain from where these Trust Properties have emerged -- or which properties were excluded -- except that they were discovered ``during research'' and either passed or failed ``logic tests''. Which and whose research, and what logic, we do not know. One could say that we are being asked to \textit{blindly trust} the methodology, which OSSTMM, ironically, tells us should be avoided. While we might accept that a property such as \textit{size} is addable and comprises quantifiable elements, other properties hold no calculable features -- claiming that they do renders the methodology increasingly futile. To this end, let us take a closer look at one of the Trust Properties identified by OSSTMM\@: \textit{consistency} -- which is defined as the ``historical evidence of compromise or corruption of the target''~\cite[Ch.5,~p.90]{Herzog10}.

\subsubsection{Rules.}\label{sec:rules} In order to make this Trust Property (like all ten Trust Properties) calculable, OSSTMM introduces Trust Rules~\cite[Ch.5,~p.91]{Herzog10}. This, it claims, translates properties into rules through a series of questions which will produce unbiased numbers as answers.  

\begin{quote}
Using the trust properties allows us to create only quantifiable rules, not ``soft'' rules that can either substantiate the trust level nor disrupt it with a biased, emotional weight. However, the properties on their own are useless if they cannot become quantifiable properties, objective, or understandable by the common person not necessarily involved in the security field.~\cite[Ch.5,~p.91]{Herzog10}
\end{quote}

OSSTMM gives an example which concerns making better hiring decisions. Here, humans (as the potential new hires) are the target, meaning that a series of human qualities are assessed; framed within the ten Trust Properties and measured using the Trust Rules. Thus, let us return to the Trust Property \textit{consistency} to explore how OSSTMM translates this into a Trust Rule that can be used in the hiring of new staff: \osstmmquote{5. \emph{Consistency}:\\
5.1. The total number of months which the applicant has not been employed divided by the total number of months the applicant has been on the workforce and eligible for employment.\\
5.2. The total number of criminal offenses known divided by the current age less eighteen years (or
the legal age of an adult in your region) of the applicant.\\
5.3. The number of neutral or negative references from past employers divided by the total
number of past employers.\\
5.4. Record the average of these results.}{\cite[Ch.5,~p.93]{Herzog10}}

These questions introduce several uncertainties and unknowns. For example, calculating the number of criminal offences ``known'' naturally ignores potentially unknown criminal offences, but more importantly, it ignores what the offences were, when they happened and how they might influence the work of the applicant.\footnote{Many legal systems distinguish between criminal convictions such as felonies and misdemeanours. Similarly, the nature of an offence is taken into account by e.g.~the Solictors Regulation Authority in England and Wales \url{https://www.sra.org.uk/solicitors/handbook/suitabilitytest/part2/content.page}.}
Moreover, assessing whether a reference is ``neutral'' or ``negative'' requires individual interpretation and qualitative judgement, which is not accounted for in OSSTMM. So, while these calculations \textit{can} be done -- i.e.~they are feasible -- and will result in a number, what this number tells us is unclear at best. Similarly, every object of calculation is chosen to be weighted exactly the same, one neutral reference per past employer and one criminal offence per year are treated the same and carry the same value. Finally, why the average number, instead of, say, the median or the max, of the answers to these questions is equally obscure.\footnote{For example, taking the computer science perspective of trust, one would expect the max i.e.~worst-case security, not average-case: one vulnerability suffices to subvert security goals.}

In summary, in OSSTMM's Trust Properties and Rules, the category mistake we have encountered when discussing the rav, i.e.~assigning -- without justification -- quantities to qualitative data, reappears.

\subsection{Risks \& Threats}

As discussed in the introduction, the reason OSSTMM gives for introducing its trust metric is to avoid risk analysis, which OSSTMM maintains ``speculates'' and ``derives opinions''~\cite[Ch.3,~p.53]{Herzog10}: 

\begin{quote}
The fundamental difference between doing a risk analysis versus a security analysis is that in security analysis you never analyze the threat.~\cite[Ch.5.~p.53]{Herzog10}
\end{quote}

Threats, in any form or shape, do therefore not form part of Actual Security. This is perhaps not surprising given OSSTMM's sole focus on what can be observed and the assertion that the existence, timing and direction of a threat can only be assumed and not known.

In OSSTMM terms, then, doing away with considerations of threats allows for the creation of ``Perfect Security, the exact balance of security and controls with operations and limitations''~\cite[Ch.1,~p.20]{Herzog10}. This invites the question of what standard it has to judge this balance as perfect. OSSTMM posits a balance of controls (adding security) with operations (taking away security) or a balance of \emph{means} to achieve security in light of functionality requirements, without discussing the \emph{end} that is pursued by these means, i.e.~against what security ought to be achieved. OSSTMM's balance is not a relation between the object and the environment but exists purely within the object by definition: 

\begin{quote}
In reality, ``perfect'' is a subjective concept and what may not be perfect for one person may indeed be perfect for another. Within the context of this manual, ``perfect'' means a perfectly balanced equation when calculating the attack surface consisting of OpSec and Limitations against Controls.~\cite[Ch.3,~p.55]{Herzog10}
\end{quote}

The balanced equation is attained when each access has all ten, unlimited controls since, as we noted above, ``having all 10 controls for each pore is functionally the same as closing the pore provided the controls have no limitations''~\cite[Ch.4,~p.67]{Herzog10}. Thus, in any field, OSSTMM's security goal is achieved when no conceivable threat has any access to the asset under protection, all pores are ``functionally'' closed. OSSTMM's security is perfectly balanced when all controls that OSSTMM can think of are in place without limitations, i.e.~when no degree of freedom exists beyond these totally encompassing controls.\footnote{Note that this is a worst-case notion of security, in contrast to the average-case notion used when applying the Trust Rules.}
In Data Network Security this approach may lead to ``interactions within a scope as well as complexity and maintenance issues''~\cite[Ch.1,~p.22]{Herzog10}. In Human Security, it develops a whole different impact to be reckoned with.

\section{Human Security}\label{sec:human-security}

Recall that one of the key features of OSSTMM is that it employs one single security testing methodology for all five channels identified in the document: Human (Chapter~7), Physical (Chapter~8), Wireless (Chapter~9), Telecommunications (Chapter~10), and Data Network Security (Chapter~11). Thus, it produces a methodology that can, in principle, be applied to anything or anyone amongst those irrespective of the object under consideration. To do this, the same 17 ``Testing Modules''\footnote{Posture Review, Logistics, Active Detection Verification, Visibility Audit, Access Verification, Trust Verification, Controls Verification, Process Verification, Training Verification, Configuration Verification, Property Verification, Segregation Review, Exposure Verification, Competitive Intelligence Scouting, Quarantine Verification, Privileges Audit, Survivability Validation, and Alert and Log Review~\cite[Ch.6]{Herzog10}.} are employed for each of the five channels; meaning that the methodology determines which aspects of each channel matter in security terms. As a result, and as we highlight in the introduction, OSSTMM considers itself  a ``holistic'' methodology and has been recognised for its ``comprehensive'' approach to security testing~\cite{CREST:Guide17}. Particularly its attention to Human Security separates it from other methodologies (see Section~\ref{sec:related-work}).

OSSTMM's treatment of trust gives a first account of how it conceptualises human interactions. For example, in the hiring process of new employees, OSSTMM suggests the following Trust Rule under the heading ``Porosity'':
 
\begin{quote}
The number of employees living in the same community as the applicant divided by the total number [of] people in the community.~\cite[Ch.5,~p.93]{Herzog10}
\end{quote}

An employee at age 50 with one prior conviction for fraud is the same security liability  (1/32) as an employee living in a town of 5,000 people together with 156 other employees. That is, an employee living in a small town with many other employees of the same organisation is considered a possible ``hole in [the] wall'', i.e.~a security threat. This might seem counter-intuitive, when starting from a perspective of, say, social cohesion, but is consequential in OSSTMM's logic. These small-town employees have social relations with other employees outside the control of the employer, i.e.~they engage in processes outside the control of it. In OSSTMM's perspective on human agency and relations, they appear as potential threats. It thus laments that social norms prevent the security enthusiast from treating people accordingly: 

\begin{quote}
Unfortunately, while using more controls works with objects and processes, it may not work between people. Many times social norms consider controls beyond simple authentication like matching a face or voice with an identity to be offensive to the person to be trusted. Society often requires us to be more trusting as individuals in order to benefit society as a whole and sometimes at the expense of everyone’s individual protection.~\cite[Ch.5,~p.87]{Herzog10}
\end{quote}

When OSSTMM rhetorically takes the standpoint of the individual's protection that is being undermined by ``society as a whole'', it practically takes the standpoint of an uninhibited authority against the individuals under its command. To this imagined authority any moment merely out of reach of surveillance and control is to the detriment of an organisation's security, which, we recall, makes no reference to threats. Using the same example as above, OSSTMM demonstrates this in another Trust Rule: 

\begin{quote}
The number of hours per day the applicant will be working alone, unassisted, unmonitored divided by the number of working hours.~\cite[Ch.5,~p.92]{Herzog10}
\end{quote}

This view of human agency, as something that needs to be controlled, is also evident throughout Chapter~7 on Human Security Testing. Here, OSSTMM takes a more pro-active approach and, for example, suggests to test Trust as: 

\begin{quote}
In Terrorem. Test and document the depth of requirements to incite fear, revolt, violence, and chaos, through the disruption of personnel and the use of rumor or other psychological abuse.~\cite[Ch.7,~p.110]{Herzog10}
\end{quote}

How OSSTMM proposes to carry out these tests is unaccounted for. How it can be done without breaking ethical guidelines and legal frameworks remains unanswered.

OSSTMM's proposition to target people through psychological means, including ``fear'', ``rumor'', and ``abuse'' to test the level to which they can be trusted is reminiscent of modern counterinsurgency operations, which, compared with traditional military campaigns, rely as much on psychological means as on physical action. A quick search through a few military counterinsurgency field manuals and doctrines, such as~\cite{MOD:Army09,USAUSM10,MOD15}, demonstrates these similarities by identifying the need to influence individual perceptions through ``aggressive'' information operations~\cite[p.152]{USAUSM10} aiming ``to influence, disrupt, corrupt, or usurp the decision making TAs [target audiences] to create a desired effect to support achievement of an objective''~\cite[p.x]{DOD14}.

This standpoint of requiring counterinsurgency-like techniques to test the resilience of an organisation paired with a desire for total control over human (inter)actions is no accident but the consequence of two of OSSTMM's key tenets: first, Human Security is treated identically to all other areas of security; humans are treated as objects, just like computers, buildings and so on. Second, security is defined without any regard to risk or threat.

While, as discussed above, perfect security is conceptualised as a balance of operations and controls, this balance itself has no reference to any threat. Hence, OSSTMM contains no notion of proportionality; whether a control measure is justified in light of a threat or not cannot be determined given OSSTMM's rejection of threats altogether. To illustrate this consider the legal category of proportionality first developed by High State Administrative Courts in Germany to review actions of the police, i.e.~the Security State~\cite{Hirschberg81}.\footnote{Also in military campaigns, from where OSSTMM appears to borrow some of its language and approach, proportionality is a legal obligation and according to one US Field Manual``requires that the advantage gained by a military operation not be exceeded by the collateral harm''~\cite[p.247]{USAUSM10}.} It is part of European Human Rights law under the European Convention of Human Rights (ECHR) and in UK Human Rights law it is interpreted as follows: ``it is necessary to determine (1) whether the objective of the measure is sufficiently important to justify the limitation of a protected right, (2) whether the measure is rationally connected to the objective, (3) whether a less intrusive measure could have been used without unacceptably compromising the achievement of the objective, and (4) whether, balancing the severity of the measure's effects on the rights of the persons to whom it applies against the importance of the objective, to the extent that the measure will contribute to its achievement, the former outweighs the latter''~\cite{UKSC39}.

In contrast, OSSTMM, while referencing human rights as a consideration ``to assure a safe, high quality test''~\cite[Ch.7,~p.105]{Herzog10}, is conceptually incapable of ``balancing the severity of the measure's effects on [the rights of] the persons to whom it applies against the importance of the objective''.\footnote{This is not a claim about OSSTMM's legality but ought to explain that its heavy-handed approach to human security is a logical consequence of its conception of security.}

OSSTMM does not see people as a strength in security terms but holds an increasingly criticised view of humans in the security loop (see e.g.~\cite{HSE:PSF14}), one where humans are always perceived as the ``weakest link'' rather than as the subject of security. Instead, OSSTMM anticipates that people will ``make mistakes, forget tasks, and purposefully abandon tasks''~\cite[Ch.2,~p.33]{Herzog10}, and it aims to eradicate these ``traits'' by reducing the human to an object whose interactions should be monitored, controlled and restricted. It does so by developing a methodology that replaces the notion of people as subjects of security with one where people are understood solely as objects of security. It models human agency as a wild-card to be controlled: 

\begin{quote}
Unfortunately, how we interact is just based on a collection of biases we accumulate during life, which are subjected to the emotional or bio-chemical state we are under when we have them.~\cite[Ch.14,~p.204]{Herzog10}
\end{quote}

This position contradicts most writings in security studies which recognise the human as a critical security actor, e.g.~\cite{BWD98} and explored further below in Section~\ref{sec:related-work}. The rejection of human agency is not only a problem in academic terms. By not seeing the human as a key security actor, capable of doing the securing, security itself is weakened. OSSTMM chooses to frame human agency as a problem, rather than as a potential solution. This view does not stop with those being tested. Indeed, OSSTMM's suspicion of human agency also applies to those doing the testing:
 
\begin{quote}
We are, after all, only human. Most often though our opinions are limited and restricted to a small scope we know as ``our little world''. We apply them everywhere because they make life easier. But when we take them with us and try to adhere them to larger, different, more complicated series and types of interactions, we will likely make mistakes. What may make perfect sense to us based on our experiences may not make any sense at all outside of ``our little world''. So what we need is a better, less biased way of looking at the bigger, more dynamic, less personal, world beyond ourselves.~\cite[Ch.14,~p.204]{Herzog10}
\end{quote}

OSSTMM's methodology for studying human security eliminates the human from security: both in the form of the recognising subject -- the tester -- whose verdicts it fundamentally distrusts as biased and replaces with a score with little meaning, and in the form of recognised subjects -- the tested -- whose agency is a ``hole in [the] wall''. In OSSTMM's view, the human mind is a security threat.

\section{Discussion}\label{sec:conclusion}

We have reviewed OSSTMM's main contributions to the field of security testing. In Section~\ref{sec:the-rav}, we observed that OSSTMM commits a type error by treating categorical values that are specific to different domains as ordinal values that apply across a wide range of domains. As a consequence it presents a security score using homographs and unmotivated choices which has little relation to the object being measured. In Section~\ref{sec:value-of-trust}, we observed that OSSTMM's notion of Trust confuses and identifies different notions of trust, producing nonsensical claims. This lack of clarity serves to create a unified Trust score to replace considerations of risk and threats, producing a notion of security that is internal to the object being studied, independent of attacker goals or capabilities. Finally, in Section~\ref{sec:human-security}, we criticised the effects of OSSTMM's approach to identify Human Security with other areas of security, disregarding the subjectivity of the objects under consideration. This produces an approach to security testing that alienates those who are relied on to do the securing, which ultimately weakens security itself.

In summary, we found none of OSSTMM's key contributions to survive under scrutiny and that the flaws identified in OSSTMM render it incorrect. These flaws, however, are an artefact of OSSTMM's ambition to be ``scientific'', which it characterises as ``not [being] about believing or relying on your experience, no matter how vast, but on knowing facts we can build upon''~\cite[Ch.3,~p.53]{Herzog10}. In OSSTMM's view this means quantifying its data.\footnote{``It appears that we are capable to rationalize in a way to supersede how we feel about trusting a target. This means we can quantify it by applying a logical process''~\cite[Ch.5,~p.87]{Herzog10}} As we have shown, to square this circle OSSTMM has to rely on unmotivated choices and assertions throughout. OSSTMM does not provide justification or evidence to ground its methodology, rendering it unscientific on its own terms. We conclude that the serious flaws identified in our analysis make the methodology futile. Thus, we suggest that security professionals abandon the use of OSSTMM as a guide to security testing.

\begin{remark}
It might be objected that OSSTMM has utility despite the flaws identified in this work. Indeed, in addition to its conceptualisation of security, it -- like any other security testing guideline -- also does advise a security tester to scan, say, port 80 on each host in a network using a TCP SYN scan, which is sound advice. We remark, however, that the flaws we identified invalidate OSSTMM's key concepts, as expressed by ISECOM itself and others, i.e.~OSSTMM in its own right. Put differently, removing what makes OSSTMM OSSTMM from OSSTMM might result in a functional, albeit by now somewhat outdated, security testing checklist.
\end{remark}

\subsection{Related Work}\label{sec:related-work}

OSSTMM's claim that it is a ``holistic'' security testing methodology that covers both technological and human security, with a scoring system that captures trust as well as access and visibility, necessitates engagement with a diverse set of literatures and bodies of work. 

\subsubsection{Security Metrics.} A growing number of works focus on the use of security metrics and the benefits of such metrics to organisations, e.g.~\cite{KovHal06,Payne06,Jaquith07,Hayden10,CST:BreHud11,NS:Brazil14,Campbell14}, while only a small body of writings offers critical reflections on such claims and on the wide use of metrics to satisfy security assessments. For example, in an overview of security metrics Ahmed~\cite[p.108]{CTT:Ahmed16} writes: ``They [security metrics] do not provide any help in measuring or monitoring the effectiveness of controls. Instead they measure the existence of controls''. Moreover, Kaur and Jones~\cite[p.45]{KauJon08} note the tension embedded in security metrics: ``It is difficult to have one metrics [sic.] that covers all types of devices. To be effective the level of detail and granularity needed is high. However, to have a large scope and cover all manner of devices requires a general metrics [sic.] which will not meet all security challenges''. Recently, CERT/CC published a critique of the widely used Common Vulnerability Scoring System (CVSS)~\cite{SHHMS18} where the authors note the data type error committed by CVSS by treating ordinal values as interval values, the difficulty of assigning a single score irrespective of differing requirements and suggest ``the way to fix this problem is to skip converting qualitative measurements to numbers''. Hence, while security metrics and standardised forms of assessing the robustness of an organisation's security posture are widely used and generally accepted, some criticisms, albeit a small selection, do exist.

\subsubsection{Metrics.} Beyond information security, the pitfalls of translating qualitative statements into quantities (which may then be algebraically manipulated) are a subject of active debate in the social, psychological and clinical sciences, see e.g.~\cite{KuzUrbMcc96,PorSal13} and the references therein. These discussions testify to the need for critical engagement with the methods of evaluation before offering conclusive findings. This is also evident in economics, where e.g.~Kay criticises the ``modern curse of bogus quantification''~\cite{Kay11} and points out that the ``index [\ldots] is not telling us anything we have not already told the index [\ldots]''~\cite{Kay09}. Such critiques thus warn against relying on metrics for a complete understanding of the objects under evaluation.   

\subsubsection{Trust.} Measuring trust is a key component of OSSTMM\@. Yet, as evidenced in our critique and exemplified by a broad range of literature, trust is not a monolithic concept. It carries multiple meanings depending on perspective, purpose and disciplinary grounding. Within the social and behavioural sciences, disciplines such as sociology, (social) psychology and behavioural economics disagree on trust definitions and methodological approaches, see e.g.~\cite{BenHal10,Cook05,CoLeHa09,Kramer99,RotSto08,Hardin04,Hardin13}, which in turn differ from notions of trust in computer science, e.g.~\cite{BlFeJo96,LeeNas10,GliWin11}. A common understanding of trust does therefore not exist. This is epitomised by Mollering~\cite{Mollering06} when outlining three schools of thought on trust: trust as reason, trust as routine and trust as reflexive. More specifically, for (social) psychologists, e.g.~\cite{BieVor04,Dunning11}, the notion of trust as a rational choice -- as a decision -- dominates, whilst behavioural economic understandings of trust focus on averting negative outcomes of opportunism and to limit the risk of exploitation~\cite{Lindenberg00,Tullberg08}. These positions rely heavily on individualistic and psychological positions put forward by experimental and quantitative researchers, and they contrast with the notion of trust as a ``sociological reality'' where ``[t]rust in everyday life is a mix of feelings and rational thinking''~\cite{LewWei85}. While sociology defines trust in relation to social processes and relations, behavioural notions of trust founded in psychology or behavioural economics take an individualistic approach.

While some seminal writings have placed trust at the centre of sociological theorising~\cite{Luhmann79,Barber83}, most computer security perspectives on humans largely ignore this branch of trust research -- trust as a social construction -- and, instead, conceptualise trust in line with psychology and behavioural economics~\cite{LeeNas10,BlFeJo96,Colwill09} as their primary aim is usability, e.g.~\cite{BlyCovLit15,CJLHWB13,KOBKSJ08,RieSasMcc03,RieSasMcc05,RieSasMcc07,Taddeo10}. This therefore also leads to a reductive and individualistic view of trust, and the mistaken assumption that trust and~\emph{trusted} in computer science carry the same meaning as in the social sciences. For example, Camp et al.~\cite[p.96]{CaNiMc01} note: ``trusted in the social sciences has exactly the same meaning of trusted in computer science [\ldots] that which is trusted is trusted exactly because if it fails there is a loss''. This not only assumes a common understanding of ``trusted'' in social and computer science, but also within the social sciences themselves; an assumption that is invalidated by the works cited above.
From this non-exhaustive, yet, multi-perspective discussion on trust, it is evident that no single definition exists and that each distinct definition of trust serves its own specific purpose.

\subsubsection{Human Agency.} As we have seen, applying its security tests to humans is a key aspect of OSSTMM; an aspect that also distinguishes it from other security testing methodologies, see e.g.~\cite{PraRam10}. Critically, however, humans have agency which means that they have the capacity to act independently or collectively upon their environment, to influence their surroundings and to make choices. The notion of human agency has also received increased attention in scholarly writings pertaining to organisational consequences of technological advancements, e.g.~\cite{ASQ:FelPen03,ISR:SchOrl04,OS:BouRob05}. In this body of literature, the notion of an ``agentic turn'' describes the increased agency of actors in relation to the organisation. From this perspective, security processes and technologies are shaped as much by the humans that use them as by their material objects.

However, Pfleeger and Caputo~\cite{CS:PflCap12} note that while a key aspect of improving information security involves understanding human behaviour, most efforts ``focus primarily on incorporating new technological approaches in products and processes''. Similarly, Sasse and Flechais~\cite{SasFle05} argue that a secure system is a sociotechnical system based on an understanding of human behaviours to ``prevent users from being the `weakest link'\,''. This is a view that was already cemented in Adams and Sasse's seminal work. It showed that a lack of understanding of users resulted in an absence of user-centred design in security mechanisms~\cite{AdaSas99}, which led users to both intentionally and unintentionally circumvent such mechanisms (see also later work, e.g.~\cite{CJLHWB13,FurCla12,Woltjer17}). Such writings evidence the critical need to recognise and understand human agency in security terms, rather than treating humans as passive objects that need security to be done~\emph{to} them.

\subsubsection{Security Testing Methodologies.} OSSTMM is one of a handful of established penetration testing methodologies, standards and guidances such as~\cite{RBDHBSRC06,NIST:SP800-115,Council15,PTES,MeuMul14}. Most of these documents focus on technical steps to be carried out by the tester~\cite{RBDHBSRC06,MeuMul14}, while some focus more on pre- and post-engagement~\cite{Council15} or provide a combination of both~\cite{PTES}. Besides OSSTMM no methodology lays any claim to being scientific or attempts to capture such a broad range of areas in which security could be considered.\footnote{``Therefore, with version 3, the OSSTMM encompasses tests from all channels -- Human, Physical, Wireless, Telecommunications, and Data Networks. This also makes it a perfectly suited for testing cloud computing, virtual infrastructures, messaging middleware, mobile communication infrastructures, high-security locations, human resources, trusted computing, and any logical processes which all cover multiple channels and require a different kind of security test.''~\cite[Introduction,~p.11]{Herzog10}} Indeed, typically these methodologies focus on network- and infrastructure penetration testing, while specialised methodologies for web applications exist~\cite{MeuMul14}, i.e.~the focus in other documents is considerably narrower than in OSSTMM\@. These methodologies are largely compatible by exhibiting a significant level of similarity in suggesting variants of a stepped discovery, enumeration and exploitation approach. It is worth noting, however, that regardless of this similarity some methodologies are incompatible. For example, the popular Penetration Testing Execution Standard~\cite{PTES} and OSSTMM are incompatible. The former prominently features threat modelling, whereas the latter insists on disregarding threats.

Academic treatments of OSSTMM or penetration testing methodologies as objects of study only come in the form of comparisons of various methodologies, either in the preliminaries of academic works to justify their particular choice of methodology or as publications in their own right, e.g.~\cite{Shackleford14,ShaJoh15,KCSK15}. However, these comparisons restrict their attention to high-level features, such as the level of detail or what is and is not covered, as well as the genealogies of the different versions. To our knowledge, prior to this work no work existed in the literature that conceptually examines penetration testing methodologies on whether they deliver on what they promise.

\subsection{Future Work}\label{sec:future-work}

This work provokes the question of whether broader lessons can be drawn from it. While OSSTMM might be unsuitable for interrogating security, do its failings point to broader issues that should be addressed?

On the one hand, other standard security testing methodologies, such as PTES~\cite{PTES} or OWASP~\cite{MeuMul14}, avoid many of the issues which we criticise in this work. They do not define scores, they do not posit new notions of trust and they do not focus on human and social aspects of security. On the other hand, some of the issues we highlight in OSSTMM are more general.

\subsubsection{Scores.} While OSSTMM expresses the methodological dogma that scientific knowledge equals quantification particularly crudely this is not its privilege.\footnote{Perhaps the most prominent quote in this spirit is attributed to Lord Kelvin: ``When you can measure what you are speaking about, and express it in numbers, you know something about it; when you cannot measure it, when you cannot express it in numbers, your knowledge is of a meager and unsatisfactory kind; it may be the beginning of knowledge, but you have scarcely, in your thoughts, advanced to the stage of science.''} Rather, this conviction is common across information security, as exemplified, for example, in CVSS which claims to score security vulnerabilities by a single magnitude. Moreover, the somewhat bad reputation of security testing as a ``tickbox exercise'' speaks of the same limitation: counting rather than understanding. Echoing the critique of CVSS in~\cite{SHHMS18}, we thus suggest, too, that security professionals ``skip converting qualitative measurements to numbers''. The healthy debates in other disciplines (see Section~\ref{sec:related-work}) provide material for a debate within information security to examine the correctness and utility of assigning numerical values to various pieces of data.

\subsubsection{Social.} A mistake we criticise in OSSTMM is the failure to recognise that the moments of a social organisation are different from the moments of a computer network. This, too, is no privilege of OSSTMM as can be easily verified by the prevalence of mantras along the lines of ``humans/people/users are the weakest link''. This standpoint, which is as prevalent as it is wrong~\cite{AdaSas99,HSE:PSF14}, offers the curious indictment that people fail to integrate into a piece of technology that does not work for them. In the context of security testing this standpoint has a home under the heading of ``social engineering'' and its most visible expression: routine but ineffective phishing simulations~\cite{KirSas12}. It is worth noting, though, that even when the focus is exclusively on technology, not engaging with the social relations that this technology ought to serve may produce undesirable results, for example leading to designs of technological controls with draconian effects where less invasive means would have been adequate~\cite{DanGue10}.

More broadly, the tendency of information security to rely on psychology, dominated by individualistic and behavioural perspectives and quantitative approaches to understanding social and human aspects of security~\cite{BlyCovLit15,CJLHWB13,FurCla12}, may represent an obstacle. Alternative methodological approaches from the social sciences, particularly from sociology and even anthropology, such as semi-structured interviews, participant-led focus groups and ethnography offer promising avenues to deeply understand the security practices and needs in an organisation, see e.g.~\cite{GreRenFlo15}.

\bibliographystyle{alpha}
\bibliography{local}

\appendix

\section{A Synthetic Toy Example}\label{sec:toy-example}

In the following, we symbolically work through OSSTMM's calculation~\cite[Section~4.4]{Herzog10} of the rav using a synthetic toy example of one host in a network providing a remote login service\@. To keep expressions compact we also do not model all the accesses, trusts and controls typically found in such a scenario. OSSTMM starts from the scope's Visibility \(P_{V}\), Access \(P_{A}\) and Trust \(P_{T}\). In our case, one host is visible \(P_{V} = 1h\) which responds on one port \(P_{A} = 1p\) (and on the IP layer, which we do not model here). In the interest of compactness, we will also assume \(P_{T} = 0\). OSSTMM defines Operational Security as \(OpSec_{sum} = P_{V} + P_{A}+ P_{T} = 1h + 1p\). From this, OSSTMM computes the the Operational Security base value \[OpSec_{base} = \log^{2}\left(1 + 100h + 100p\right).\] Considering a logarithm is motivated in~\cite[Section~1.5]{Herzog10} and~\cite[Section~4.1]{Herzog10} but the particular choice of \(\log^{2}(\cdot)\) is not motivated, the additive factor \(1\) is motivated to obtain zero in case of no attack surface. OSSTMM defines control meta classes \(A = \{Au, Id, Re, Su, Ct\}\) and \(B = \{NR, It, Pr, Cf, Al\}\), see Section~\ref{sec:the-rav}. We assume only authentication \(LC_{Au}\) is enforced via a login \(LC_{Au} = 1\ell\). In OSSTMM's terms the sum of loss controls in this example is thus \[LC_{sum} \sum_{\lambda \in A \cup B} LC_{\lambda} = 1\ell.\] Let \(\lambda \in A \cup B\) be any control type. Then, OSSTMM further defines Missing Controls as \[MC_{\lambda} = \max\left(OpSec_{sum} - LC_{\lambda}, 0 \right)\] and the sum of these missing controls as \(MC_{sum} = \sum_{\lambda \in A \cup B} MC_{\lambda}\). True Controls are then defined as
\begin{align*}
  TC_{\lambda} &= OpSec_{sum} - TC_{\lambda} = OpSec_{sum} - \max\left(OpSec_{sum} - LC_{\lambda}, 0 \right)\\
               &= \min\left(LC_{\lambda}, OpSec_{sum}\right) 
\end{align*}
In our case, we would have to decide if \(1h + 1p > 1\ell\), i.e.~if one active host added to one port is greater than a login being applied. OSSTMM considers all formal variables as equal to the integer 1, suggesting the inequality does not hold in OSSTMM's model. The idea of this check is you cannot miss less than zero controls. OSSTMM does not motivate why controls are not normalised by \(1/10\) here in contrast to other formulas. Following along, we obtain the True Controls base \(TC_{base}\) and the Full Controls base \(FC_{base}\):
\begin{align*}
  TC_{base} &= \log^{2}\left(1 + 100 \times \left(OpSec_{sum} - MC_{sum}/10 \right)\right)\\
            &= \log^{2}\left(10\ell + 1\right) \\
            & =\log^{2}\left(1 + 10\times LC_{sum}\right) \\
            &= FC_{base}
\end{align*}

\begin{figure*}
  \begin{tiny}
\begin{align*}
  SecLim_{sum} &= (L_V \times (OpSec_{sum} + MC_{sum})/OpSec_{sum}) + (L_W \times (OpSec_{sum}   + MC_A)/OpSec_{sum}) \\
               &+ (L_C \times (OpSec_{sum}   + MC_B)/OpSec_{sum}) \\
               &+ (L_E \times ((P_V+P_A)\times MC_{vg} + L_V + L_W + L_C)/OpSec_{sum}) \\
               &+ (L_A \times (P_T\times MC_{vg} + L_V + L_W + L_C)/OpSec_{sum})\\
               &= \frac{{\left(11 \, h - \ell + 11 \, p\right)}^{2}}{{\left(h + p\right)}^{2}}
               + \frac{{\left(10 \, h - \ell + 10 \, p + \frac{10 \, {\left(11 \, h - \ell + 11 \, p\right)}}{h + p} + \frac{10 \, {\left(6 \, h - \ell + 6 \, p\right)}}{h + p} + 60\right)}^{2}}{100 \, {\left(h + p\right)}^{2}}\\
               &+ \frac{{\left(6 \, h - \ell + 6 \, p\right)}^{2}}{{\left(h + p\right)}^{2}}
                + \frac{{\left(\frac{11 \, h - \ell + 11 \, p}{h + p} + \frac{6 \, h - \ell + 6 \, p}{h + p} + 6\right)}^{2}}{{\left(h + p\right)}^{2}} + 36
\end{align*}
\end{tiny}
\caption{Security Limitations}\label{fig:seclim}
\end{figure*}

These expressions are then combined into various Limitations Formula encoding vulnerabilities \(L_{V}\), weaknesses \(L_{W}\), concerns \(L_{C}\), exposures \(L_{E}\), anomalies \(L_{A}\) which are then combined to obtain the Security Limitations sum as given Figure~\ref{fig:seclim} where \(MC_{vg} = MC_{sum}/(10\times OpSec_{sum})\). As before, \(SecLim_{base}\) is defined as \(\log^{2}(1 + 100 \times SecLim_{sum})\). Finally, Actual Security, the ``true state of security as a hash of all three sections''~\cite[p.85]{Herzog10} is defined as
\begin{align*}
  ActSec &= 100 + FC_{base} - OpSec_{base} - \log(1 + 100 \times SecLim_{sum}) \\
         & - 1/100 \times OpSec_{base} \times (FC_{base} - \log(1 + 100 \times SecLim_{sum}))\\
         & + 1/100 \times FC_{base} \times \log(1 + 100 \times SecLim_{sum}).
\end{align*}
Expanding this formula for our toy example produces Figure~\ref{fig:actsec}. To compute the numerical rav value OSSTMM evaluates this expression at one for all formal variables. In our example, this gives a value of \(\approx -12\). The reader is invited to compare information provided by the symbolic or numerical Actual Security to the information provided by our initial informal description of the toy example. We stress that while the methodology, for example, forces the analyst to recognise that our login service provides no confidentiality (\(LC_{Cf} = 0\) in OSSTMM terms), this does not distinguish OSSTMM from other security testing methodologies. Rather, OSSTMM's key procedure is the rav computation producing Figure~\ref{fig:actsec}.

\begin{figure*}
  \begin{tiny}
  \begin{dmath*}
    ActSec = \log\left(\frac{19401 \, h^{4}}{h^{4} + 4 \, h^{3} p + 6 \, h^{2} p^{2} + 4 \, h p^{3} + p^{4}} - \frac{3420 \, h^{3} {\ell}}{h^{4} + 4 \, h^{3} p + 6 \, h^{2} p^{2} + 4 \, h p^{3} + p^{4}} + \frac{201 \, h^{2} {\ell}^{2}}{h^{4} + 4 \, h^{3} p + 6 \, h^{2} p^{2} + 4 \, h p^{3} + p^{4}} + \frac{77604 \, h^{3} p}{h^{4} + 4 \, h^{3} p + 6 \, h^{2} p^{2} + 4 \, h p^{3} + p^{4}} - \frac{10260 \, h^{2} {\ell} p}{h^{4} + 4 \, h^{3} p + 6 \, h^{2} p^{2} + 4 \, h p^{3} + p^{4}} + \frac{402 \, h {\ell}^{2} p}{h^{4} + 4 \, h^{3} p + 6 \, h^{2} p^{2} + 4 \, h p^{3} + p^{4}} + \frac{116406 \, h^{2} p^{2}}{h^{4} + 4 \, h^{3} p + 6 \, h^{2} p^{2} + 4 \, h p^{3} + p^{4}} - \frac{10260 \, h {\ell} p^{2}}{h^{4} + 4 \, h^{3} p + 6 \, h^{2} p^{2} + 4 \, h p^{3} + p^{4}} + \frac{201 \, {\ell}^{2} p^{2}}{h^{4} + 4 \, h^{3} p + 6 \, h^{2} p^{2} + 4 \, h p^{3} + p^{4}} + \frac{77604 \, h p^{3}}{h^{4} + 4 \, h^{3} p + 6 \, h^{2} p^{2} + 4 \, h p^{3} + p^{4}} - \frac{3420 \, {\ell} p^{3}}{h^{4} + 4 \, h^{3} p + 6 \, h^{2} p^{2} + 4 \, h p^{3} + p^{4}} + \frac{19401 \, p^{4}}{h^{4} + 4 \, h^{3} p + 6 \, h^{2} p^{2} + 4 \, h p^{3} + p^{4}} + \frac{4600 \, h^{3}}{h^{4} + 4 \, h^{3} p + 6 \, h^{2} p^{2} + 4 \, h p^{3} + p^{4}} - \frac{860 \, h^{2} {\ell}}{h^{4} + 4 \, h^{3} p + 6 \, h^{2} p^{2} + 4 \, h p^{3} + p^{4}} + \frac{40 \, h {\ell}^{2}}{h^{4} + 4 \, h^{3} p + 6 \, h^{2} p^{2} + 4 \, h p^{3} + p^{4}} + \frac{13800 \, h^{2} p}{h^{4} + 4 \, h^{3} p + 6 \, h^{2} p^{2} + 4 \, h p^{3} + p^{4}} - \frac{1720 \, h {\ell} p}{h^{4} + 4 \, h^{3} p + 6 \, h^{2} p^{2} + 4 \, h p^{3} + p^{4}} + \frac{40 \, {\ell}^{2} p}{h^{4} + 4 \, h^{3} p + 6 \, h^{2} p^{2} + 4 \, h p^{3} + p^{4}} + \frac{13800 \, h p^{2}}{h^{4} + 4 \, h^{3} p + 6 \, h^{2} p^{2} + 4 \, h p^{3} + p^{4}} - \frac{860 \, {\ell} p^{2}}{h^{4} + 4 \, h^{3} p + 6 \, h^{2} p^{2} + 4 \, h p^{3} + p^{4}} + \frac{4600 \, p^{3}}{h^{4} + 4 \, h^{3} p + 6 \, h^{2} p^{2} + 4 \, h p^{3} + p^{4}} + \frac{105800 \, h^{2}}{h^{4} + 4 \, h^{3} p + 6 \, h^{2} p^{2} + 4 \, h p^{3} + p^{4}} - \frac{18400 \, h {\ell}}{h^{4} + 4 \, h^{3} p + 6 \, h^{2} p^{2} + 4 \, h p^{3} + p^{4}} + \frac{800 \, {\ell}^{2}}{h^{4} + 4 \, h^{3} p + 6 \, h^{2} p^{2} + 4 \, h p^{3} + p^{4}} + \frac{211600 \, h p}{h^{4} + 4 \, h^{3} p + 6 \, h^{2} p^{2} + 4 \, h p^{3} + p^{4}} - \frac{18400 \, {\ell} p}{h^{4} + 4 \, h^{3} p + 6 \, h^{2} p^{2} + 4 \, h p^{3} + p^{4}} + \frac{105800 \, p^{2}}{h^{4} + 4 \, h^{3} p + 6 \, h^{2} p^{2} + 4 \, h p^{3} + p^{4}}\right)^{2} \\
    \cdot \left(\frac{1}{100} \, \log\left(100 \, h + 100 \, p + 1\right)^{2} -\frac{1}{100} \, \log\left(10 \, {\ell} + 1\right)^{2} -1 \right)\\
    -\frac{1}{100} \, {\left(\log\left(10 \, {\ell} + 1\right)^{2} + 100\right)} \log\left(100 \, h + 100 \, p + 1\right)^{2} + \log\left(10 \, {\ell} + 1\right)^{2} + 100
  \end{dmath*}
  \end{tiny}
  \caption{Actual Security}\label{fig:actsec}
\end{figure*}

\end{document}